\documentclass{aa}
\usepackage{psfig}

\newcommand{\msun}{M_{\odot}}
\newcommand{\nh}{N_{\rm H}}
\newcommand{\cmsq}{{\rm cm}^{-2}}
\newcommand{\ergs}{{\rm erg\,s}^{-1}}
\newcommand{\ergcm}{{\rm erg\,cm}^{-2}}
\newcommand{\ergcms}{{\rm erg\,cm}^{-2}{\rm s}^{-1}}

\begin{document}
\thesaurus{02.01.2, 08.14.1, 08.09.02, 13.25.1 }
\title{The longest thermonuclear X-ray burst ever observed?}
\subtitle{A BeppoSAX Wide Field Camera observation of 4U\,1735$-$44}
\author{R.\ Cornelisse\inst{1,2} \and J.\ Heise\inst{1} \and 
        E.\ Kuulkers\inst{1,2} \and F.\ Verbunt\inst{2}
        \and J.J.M.\ in 't Zand\inst{1}}
\offprints{R.\ Cornelisse}
\mail{R.Cornelisse@sron.nl}

\institute{Space Research Organization Netherlands, Sorbonnelaan 2, 
              3584 CA Utrecht, The Netherlands  
         \and  Astronomical Institute,
              P.O.Box 80000, 3508 TA Utrecht, The Netherlands 
                }

\date{\today / Accepted date}   
 
\maketitle

\begin{abstract}
A long flux enhancement, with an exponential decay time of 86\,min, is
detected in 4U\,1735$-$44 with the BeppoSAX Wide Field Cameras. We argue
that this is a type\,I X-ray burst, making it the longest such burst
ever observed. 
Current theories for thermonuclear bursts predict shorter and more frequent
bursts for the observed persistent accretion rate.
\end{abstract}

\section{Introduction}

Of the $\simeq$150 low-mass X-ray binaries known in our galaxy, about 40\%\ 
show occasional bursts of X-rays, in which a
rapid rise, lasting from less than a second to $\simeq$10\,s, is followed
by a slower decay, lasting between $\simeq$10\,s to minutes.
During the decay the characteristic temperature of the X-ray spectrum
decreases.
An X-ray burst is explained as energy release by rapid nuclear fusion of 
material on the surface of a neutron star and thus an X-ray burst is thought
to identify the compact object emitting it unambiguously as a neutron
star.
If the burst is very luminous, reaching the Eddington limit $L_{\rm Edd}$,
the energy release may temporarily
lift the neutron star atmosphere to radii of order 100\,km.
Reviews of observations of X-ray bursts are given by Lewin et al.\ (1993, 
1995).

The properties of a burst depend, according to theory, on the mass and
radius of the neutron star, on the rate with which material is accreted
onto the neutron star, and on the composition of the accreted material.
It is hoped that a detailed study of X-ray bursts can be used to determine
the mass and radius of the neutron star, via the relation between
luminosity, effective temperature and flux, and via the changes in the
general relativistic correction to  this relation when the atmosphere expands 
from the neutron star surface to a larger radius.
However, the physics of the X-ray burst is complex.
There is evidence that the emitting area does not cover the whole neutron star
and changes with the accretion rate.
Reviews of the theory of X-ray bursts are given by Bildsten (1998, 2000).
\nocite{lpt93}\nocite{lpt95}\nocite{bil00}\nocite{bil98}

In this paper we describe a long flux enhancement that we observed with the 
Wide Field Cameras of BeppoSAX in the X-ray burst source 4U\,1735$-$44,
and argue that this event is the longest type I X-ray burst ever observed.
In Sect.\,2 we describe the observations and data extraction,
in Sect.\,3 the properties of the flux enhancement.
A discussion and comparison with earlier long bursts is given in Sect.\,4. 
In the remaining part of this section we briefly describe earlier observations
of 4U\,1735$-$44.

4U\,1735$-$44 is a relatively bright low-mass X-ray binary.  
Smale et al.\ (1986) fit EXOSAT data in the 1.4-11\,keV range with 
a power law of photon index 1.8 with an exponential cutoff above 7\,keV,
absorbed by an interstellar column $\nh\simeq 5\times 10^{20}\cmsq$.
The flux in the 1.4-11\,keV range is $\simeq4\times10^{-9}\ergcms$.
Van Paradijs et al.\ (1988) show that a sum of thermal bremsstrahlung
of $\simeq10$\,keV and black body radiation of $\simeq2$\,keV, absorbed by an
interstellar column $\nh<8\times 10^{20}\cmsq$, adequately describes
EXOSAT data in the same energy range and at a similar flux level,
obtained one year later. A similar spectrum, with a higher absorption
column  $\nh\simeq 3.4\times 10^{21}\cmsq$, fits the Einstein
solid-state spectrometer and monitor proportional counter data 
(Christian \&\ Swank 1997). During GINGA observations, the source was
somewhat brighter, at $\simeq9\times10^{-9}\ergcms$ in the 1-37\,keV range
(Seon et al.\ 1997).

Bursts were detected  at irregular time intervals during each of
the five occasions in 1977 and 1978 that SAS-3 observed 4U\,1735$-$44,
leading to a total of 53 detected bursts (Lewin et al.\ 1980).
EXOSAT detected one burst in 1984 (Smale et al.\ 1986) and five bursts 
during a continous 80\,hr observation in 1985 (Van Paradijs et al.\ 1988),
one rather bright burst was detected with GINGA in 1991 (Seon et al.\ 1997),
and five X-ray bursts with RXTE in 1998 (Ford et al.\ 1998).
Burst intervals range from about 30 minutes to more than 50 hrs.
Three of the  bursts observed with EXOSAT and the single
burst observed with GINGA were radius expansion bursts (Damen et al.\ 1990,
Seon et al.\ 1997), and have been used to determine the distance
to 4U\,1735$-$44 as about 9.2\,kpc (Van Paradijs and White 1995).

4U\,1735$-$44 was the first X-ray burster for which an optical counterpart
was found: V926 Sco (McClintock et al.\ 1977). From optical
photometry an orbital period of 4.65 hrs was derived (Corbet et 
al.\ 1986). 
\nocite{lpch80}\nocite{scc+86}\nocite{ppl+88}\nocite{smy+97}\nocite{dml+90}
\nocite{pw95}\nocite{mbd+77}\nocite{ctc+86}\nocite{fkp+98}

\section{Observations and data extraction}

The Wide Field Camera experiment (Jager et al. 1997) is located on the
BeppoSAX platform which was launched early 1996 (Boella et al.\ 1997).
It comprises two identically 
designed coded-aperture multi-wire Xenon proportional counter 
detectors. The field of view of each camera is 40$\times$40 degrees 
full width to zero response, which makes it the largest of any flown 
X-ray imaging device with good angular resolution.\nocite{jmb+97} 
The angular resolution is $5'$ full width at half maximum, and the accuracy of
the source location is upward of $0.7'$, 
depending mainly on the signal-to-noise ratio. 
The photon energy range is 2-28\,keV, and the time resolution is 0.5\,ms. 
Due to the coded mask aperture the detector data 
consist of a superposition of the background and shadowgrams of multiple 
sources. To reconstruct the sky image an algorithm 
is employed which is based on cross correlation of
the detector image with the coded mask (Jager et al. 1997).

Since the fall of 1996, the Wide Field Cameras
observe the field around the Galactic Center on a 
regular basis during each fall and spring. 
The first campaign was a nine-day near-continuous 
observation from August 21 until August 30, 1996. 
About 30\% of the time, viz.\ $\simeq$35 minutes per orbit, 
is lost due to earth occultation and due to passage through the
South Atlantic Anomaly.

\section{A long X-ray flux enhancement of 4U\,1735$-$44}

\begin{figure}
\centerline{\psfig{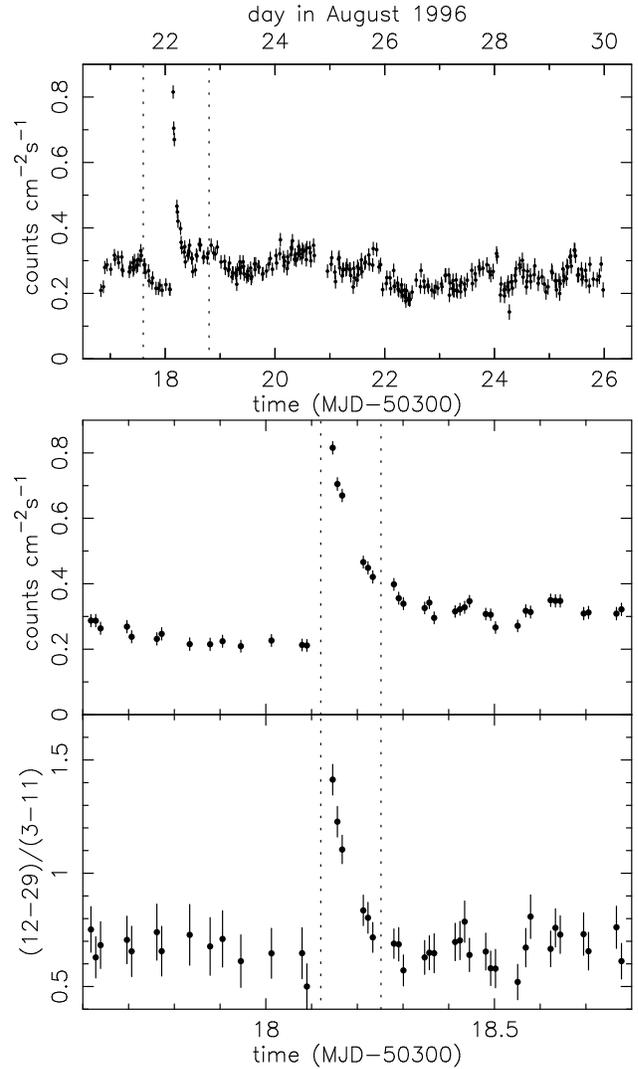}}
\caption{Top: The nine day lightcurve of 4U\,1735$-$44 as observed
with the WFC in August 1996. Countrates are for channels 1-31 (energy
range 2-28\,keV). Each time bin corresponds to
15 minutes. A large enhancement in intensity starts near MJD 50318.1 and 
ends about $\simeq$4.0 hours later. The vertical dotted lines indicate the
time interval for which the countrate and hardness ratio are shown in the 
expanded view of the lower frames.
The hardness ratio shown is the ratio of the countrate in channels
12-29 (5-20\,keV) to that in channels 3-11 (2-5\,keV).
During the flux enhancement the exponential softening expected for a type-I 
X-ray burst is clearly visible. 
The vertical dotted lines indicate the time interval for which we add the
data to obtain the burst spectrum.
\label{fcurve}}
\end{figure}

In Fig.\,\ref{fcurve} we show the lightcurve of 4U\,1735$-$44 as observed
with the WFC between 21 and 30 August 1996.
The persistent countrate varies between 0.2 and 0.3 counts\,cm$^{-2}$s$^{-1}$.
Immediately after the earth occultation on MJD 50318.1 a strong
enhancement (factor $\simeq$3) in the X-ray intensity was seen which 
subsequently decayed exponentially. An expanded lightcurve of this event
is also shown in Fig.\,\ref{fcurve}. The position derived for this event
is $3\farcs2\pm3\farcs4$ from the position of 4U\,1735$-$44 as derived
from its persistent emission.
(Both positions share the same systematic error, and thus their relative
position is much more accurate than their absolute positions, which
have errors of $\simeq 1'$.)
We conclude that the event is from 4U\,1735$-$44.

To the persistent flux we fit the two models discussed in the introduction,
i.e.\ a power law with high energy cutoff, and a sum of bremsstrahlung
and black body spectra, in the 2-24\,keV range. The spectrum before
and after the event are for the intervals MJD\,50317.8-50318.1 and
MJD\,50319.7-50320.8, respectively.
The results are given in Table\,\ref{tfit}.
We note that the values of the fit parameters are similar to those
for earlier observations. Notwithstanding the different flux levels
before and after the flux enhancement, the hardness of the spectrum (also
shown in Fig.\,\ref{fcurve}) is similar.
The persistent flux corresponds to an X-ray luminosity at 9.2\,kpc
of $4.4\times10^{37}\ergs$ in the 2-28\,keV band.
In our fits we set the interstellar absorption at a fixed value of
$\nh=3.4\times10^{21}\cmsq$; the hard energy range of the WFC is not much 
affected by absorption, and fits for different assumed absorption values
give results similar to those listed in Table~\ref{tfit}.

\begin{table}
\caption{Results of the modeling of the X-ray spectrum. We fit a
cutoff photon powerlaw spectrum 
$N(E)=N_{\rm o}E^{-\Gamma}e^{-(E-E_{\rm o})/E_{\rm w}}$
and a sum of a bremsstrahlung of temperature $T_{\rm br}$ and black body 
spectrum of temperature $T_{\rm bb}$ and radius $R$
to the data before and after the burst. For the burst, we fix the 
parameters of either the cutoff power law or the bremsstrahlung component
to the values found after the burst, and fit for a blackbody added to
these. The absorption column is fixed at the value of
$\nh=3.4\times10^{21}\cmsq$ found by Christian \&\ Swank (1997).
For each model we give the total flux
in the range observed with the WFC, i.e.\ 2-28\,keV, as well as, for
comparison with earlier observations, in the range of 1.4-11\,keV.
\label{tfit}}
\begin{tabular}{lrr}
{\bf cutoff power law} & before & after \\
$\chi_{\nu}^2$\,(dof) & 0.9\,(23) & 0.9\,(23) \\
$\Gamma$ & 1.84$\pm$0.07 & 1.55$\pm$0.06 \\
$E_{\rm o}$\,(keV) & 9.4$\pm$1.1 & 6.2$\pm$0.4 \\
$E_{\rm w}$\,(keV) & 3.6$\pm$1.7 & 6.9$\pm$0.7 \\
$F_{2-28}$\,($10^{-9}\ergcms$) & 3.44$\pm$0.10 & 4.31$\pm$0.08 \\
$F_{1.4-11}$ \,($10^{-9}\ergcms$) & 3.53$\pm$0.06 & 3.91$\pm$0.04 \\
{\bf brems plus black body} & before & after \\
$\chi_{\nu}^2$\,(dof) & 1.1\,(23) & 1.0\,(23) \\
$kT_{\rm br}$\,(keV) & 7.8$\pm$0.6 & 8.0$\pm$1.3 \\
$kT_{\rm bb}$\,(keV) & 0. & 1.6$\pm$0.2 \\
$R$\,(km)            & 0. &  3.8$\pm$0.9 \\
$F_{2-28}$ \,($10^{-9}\ergcms$) & 3.80$\pm$0.09 & 4.32$\pm$0.10 \\
$F_{1.4-11}$ \,($10^{-9}\ergcms$) & 3.40$\pm$0.05 & 3.87$\pm$0.04 \\
{\bf added blackbody for burst} & $+$cutoff & $+$brems \\
$\chi_{\nu}^2$\,(dof) & 1.3\,(25) & 1.3\,(25) \\
$kT_{\rm bb}$\,(keV) & 1.70$\pm$0.05 & 1.69$\pm$0.04 \\
$R$\,(km)            & 6.1$\pm$0.3 &  7.2$\pm$0.3 \\
$F_{2-28}$ \,($10^{-9}\ergcms$) & 7.42$\pm$0.05 & 7.43$\pm$0.05 \\
$F_{1.4-11}$ \,($10^{-9}\ergcms$) & 6.76$\pm$0.04 & 6.71$\pm$0.04 \\
\end{tabular}
\end{table}
\nocite{cs97}

To describe the flux decline we first fit an exponential $C=C(0)e^{-t/\tau}$ 
to the observed countrate in the 2-28\,keV range. 
The fit is acceptable (at $\chi^2_\nu=1.6$ for 33 d.o.f.) and 
$\tau=86\pm5$\,min.
Fits to the counts in the 2-5\,keV and 5-20\,keV ranges give decay
times of $129\pm15$ and $67\pm5$\,min, respectively, in accordance
with the observed softening of the flux during decline (see 
Fig.\,\ref{fcurve}).
We fit the spectrum during the flux enhancement as follows.
First we add all the counts obtained between MJD\,50318.10 and 50318.25.
We then fit the total spectrum with the sum of a black body and
either a cutoff power law spectrum or a thermal bremsstrahlung spectrum.
In these fits, the parameters of the power law and bremsstrahlung
component are fixed at the values obtained for the fit to the
persistent spectrum after the event.
The resulting parameters for the black body are also listed
in Table\,\ref{tfit}.
At the observed maximum the bolometric flux was 
$(1.5\pm0.1)\times 10^{-8}\ergcms$
which for a source at 9.2\,kpc corresponds to a luminosity
of $1.5\times10^{38}\ergs$.
The start of the flux enhancement is not observed, but if we assume that its
maximum is at the Eddington limit of $1.8\times10^{38}\ergs$
(for a neutron star mass of $1.4\msun$) and that the decay time
is constant, then maximum was reached  23.6 min before the source
emerged from earth occultation, leaving at most 12.4\,min for 
the rise to maximum (since the start of the data gap). 
The decay from maximum was therefore much longer,
by a factor $>$8, than the rise.
The fluence in the observed part of the burst is $5.1\times10^{-5}\ergcm$,
corresponding to $5.2\times10^{41}$\,erg; this is a lower limit to
the energy released during the full event.

We have also made fits to the first and second half of the event separately,
and find temperatures for the blackbody component of 2.1-2.2\,keV and
1.3-1.4\,keV for the first and second half respectively, confirming the
softening. For the blackbody radius we find 5.7-6.5\,km and 8.5-8.8\,km,
for the first and second half, respectively. This apparent increase in radius
is probably due to the difference between the observed colour temperature
and the actual effective temperature of the black body; when we apply
corrections to the colour temperature as given by van Paradijs et al.\ (1986)
the value for the radius in the first part of the burst increases to 14\,km,
whereas that for the second half is unchanged.\nocite{psl+86}

\section{Discussion}

In addition to the thermonuclear X-ray bursts, also called
type I bursts, low-mass X-ray binaries show other sudden enhancements
in X-ray flux. Type II bursts are different from type I bursts in that type II 
bursts do not show cooling of the characteristic temperature of the X-ray 
spectrum during the decline. X-ray flares have an irregular flux evolution.
Type II bursts are thought to be accretion events; the nature of flares
is unknown. 

The flux enhancement of 4U\,1735$-$44 shows a smooth exponential decay of the 
countrate and of the characteristic temperature. Its rise must have been 
shorter than the decline. A black body gives a good fit to the observed
spectrum, for a radius as expected from a neutron star, similar
to earlier, ordinary bursts of 4U\,1735$-$44.
All these properties indicate a type\,I burst.
The only special property of the new burst is its duration, which when
expressed as the ratio of fluence $E_{\rm b}$ and peak flux $F_{\rm max}$: 
$E_{\rm b}/F_{\rm max}>3400$\,s, is more than 300 times
longer than the longest burst observed previously from this source
(see Lewin et al.\ 1980).
This duration also translates in a fluence which is several orders of
magnitude larger than the previous record holder for 4U\,1735$-$44, because 
the peak flux is similar to those of normal type\,I bursts.
The fluence of a type\,I burst which burns all matter deposited onto a 
neutron star since the previous burst must be $\simeq$1\%\ of the accretion
energy released by deposition of this matter. We do not have a measurement
to the previous burst, but in seven days following the burst no other
burst was observed. 
Multiplying this time by the persistent luminosity we
obtain $\simeq2.7\times10^{43}$\,erg, or about 50 times the energy of the
burst, well in the range of previously observed ratios for type\,I bursts.

The presence of clear cooling argues against a type\,II burst; this and
the smooth decay argues against a flare. If the flux enhancement were due to
an accretion event, the amount of matter dropped extra onto the neutron
star (assuming a mass of $1.4\msun$ and a radius of 10\,km) must have
been $>3\times10^{21}$\,g, which may be compared to the average accretion
rate of $2.3\times10^{17}$\,g\,s$^{-1}$ derived for the persistent flux.
If the inner part of the accretion disk would have depleted
itself onto the neutron star during the flux enhancement, one would expect 
the accretion rate immediately after to be lower than before. The
observations suggest the opposite.

We conclude that a type\,I X-ray burst is the best explanation for the
enhanced flux event.
We consider it significant that the occurrence of this burst is
accompanied by the absence of any ordinary -- i.e.\ short -- burst
throughout our 9-day observation, whereas all previous observations of 
4U\,1735$-$44 did detect ordinary bursts (see Introduction).

Searching the literature for long bursts we find that the longest type\,I
burst published previously is a radius expansion burst observed with
SAS-3, probably in 4U\,1708$-$23 (Hoffman et al.\ 1978; see also
Lewin et al.\ 1995). The ratio of fluence and peak flux for that burst
was $\simeq 500$\,s, so that the BeppoSAX WFC burst of 4U\,1735$-$44 lasted 
at least six times longer.
Other events published as long bursts from Aql\,X-1 (Czerny et al.\ 1987)
and from X\,1905+000 (Chevalier and Ilovaisky 1990) are in fact relatively
short bursts followed by an enhanced constant flux level which persisted
for several hours: in both cases the flux declined to 1/e of the peak level
within 20\,s. These events are clearly different from the long exponential
bursts seen in  4U\,1708$-$23 and 4U\,1735$-$44. 
\nocite{hld+78}\nocite{ci90}\nocite{ccg87}
 
From the theoretical point of view, a long interval between bursts
would allow hydrogen to burn completely before the onset of the burst,
so that the energetics of the burst is dominated by pure helium burning.
If matter accreted at a rate of $2.3\times10^{17}$\,g\,s$^{-1}$ during one week,
the energy released by helium burning is compatible with the energy of
the observed burst. The problem with this model is that theory predicts 
for this accretion rate that the burst initiates well before hydrogen burning 
is completed, i.e.\ that bursts are more frequent and less energetic,
in accordance with those previously observed of 4U\,1735$-$44.
Indeed, Fujimoto et al.\ (1987) find that a burst of $10^4$\,s duration 
occurs only for accretion rates $\dot M<0.01\dot M_{\rm Edd}$.
The persistent flux during the BeppoSAX observation is a factor $\simeq20$
higher than this limit; observations previous
to ours have consistently found 4U\,1735$-$44 at a similar luminosity.

An alternative model for bursts with a duration of $10^4$\,s is
accretion of pure helium at an accretion rate in excess of 
the Eddington limit ($\dot M>5\times\dot M_{\rm Edd}$, Brown \&\ Bildsten
1998). 
The orbital period and optical spectrum indicate a main-sequence, 
i.e.\ hydrogen-rich, donor star (Augusteijn et al.\ 1998).
\nocite{fhir87}\nocite{bb98}

Perhaps the main challenge for any theoretical explanation is that
the properties of the persistent flux during our nine day long observation,
during which a single very long X-ray burst was observed, are not different
from those during earlier observations with EXOSAT when more frequent
ordinary bursts were found.


\begin{thebibliography}{}

\bibitem[\protect\astroncite{Augusteijn et al.}{1998}]{ahj+98}
Augusteijn, T., van der Hooft, F., de Jong, J., et al. 1998, A\&A, 332, 561
\bibitem[\protect\astroncite{Bildsten}{1998}]{bil98}
Bildsten, L. 1998,
\newblock in A. Alpar, L. Buccheri, J. van Paradijs (eds.), The many faces of
  neutron stars, NATO ASI, Kluwer, Dordrecht,  419

\bibitem[\protect\astroncite{Bildsten}{2000}]{bil00}
Bildsten, L. 2000,
\newblock in S. Holt, W. Zhang (eds.), Cosmic explosions, AIP,  in press
(astro-ph/0001135)

\bibitem[\protect\astroncite{Boella et~al.}{1997}]{bbp+97}
Boella, G., Butler, R., Perola, G., et~al. 1997, A\&AS, 122, 299

\bibitem[\protect\astroncite{Brown \& Bildsten}{1998}]{bb98}
Brown, E., Bildsten, L. 1998, ApJ, 496, 915

\bibitem[\protect\astroncite{Chevalier \& Ilovaisky}{1990}]{ci90}
Chevalier, C., Ilovaisky, S. 1990, A\&A, 228, 115

\bibitem[\protect\astroncite{Christian \& Swank}{1997}]{cs97}
Christian, D., Swank, J. 1997, ApJS, 109, 117

\bibitem[\protect\astroncite{Corbet et~al.}{1986}]{ctc+86}
Corbet, R., Thorstensen, J., Charles, P., et~al. 1986, MNRAS, 222, 15P

\bibitem[\protect\astroncite{Czerny et~al.}{1987}]{ccg87}
Czerny, M., Czerny, B., Grindlay, J. 1987, ApJ, 312, 122

\bibitem[\protect\astroncite{Damen et~al.}{1990}]{dml+90}
Damen, E., Magnier, E., Lewin, W., et~al. 1990, A\&A, 237, 103

\bibitem[\protect\astroncite{Ford et~al.}{1998}]{fkp+98}
Ford, E., van~der Klis, M., van Paradijs, J., et~al. 1998, ApJ (Letters), 508,
  L155

\bibitem[\protect\astroncite{Fujimoto et~al.}{1987}]{fhir87}
Fujimoto, M., Hanawa, T., Iben, I., Richardson, M. 1987, ApJ, 315, 198

\bibitem[\protect\astroncite{Hoffman et~al.}{1978}]{hld+78}
Hoffman, J., Lewin, W., Doty, J., et~al. 1978, ApJ (Letters), 221, L57

\bibitem[\protect\astroncite{Jager et~al.}{1997}]{jmb+97}
Jager, R., Mels, W., Brinkman, A., et~al. 1997, A\&AS, 125, 557

\bibitem[\protect\astroncite{Lewin et~al.}{1980}]{lpch80}
Lewin, W., van Paradijs, J., Cominsky, L., Holzner, S. 1980, MNRAS, 193, 15

\bibitem[\protect\astroncite{Lewin et~al.}{1993}]{lpt93}
Lewin, W., van Paradijs, J., Taam, R. 1993, Space Sci. Rev., 62, 223

\bibitem[\protect\astroncite{Lewin et~al.}{1995}]{lpt95}
Lewin, W., van Paradijs, J., Taam, R. 1995,
\newblock in W. Lewin, J. van Paradijs, E. van~den Heuvel (eds.), {X}-ray
  binaries, Cambridge U.P., Cambridge, p.~175

\bibitem[\protect\astroncite{McClintock et~al.}{1977}]{mbd+77}
McClintock, J., Bradt, H., Doxsey, R., et~al. 1977, Nat, 270, 320

\bibitem[\protect\astroncite{Seon et~al.}{1997}]{smy+97}
Seon, K., Min, K., Yoshida, K., et~al. 1997, ApJ, 479, 398

\bibitem[\protect\astroncite{Smale et~al.}{1986}]{scc+86}
Smale, A., Corbet, R., Charles, P., Menzies, J., Mack, P. 1986, MNRAS, 223, 207

\bibitem[\protect\astroncite{van Paradijs et~al.}{1988}]{ppl+88}
van Paradijs, J., Penninx, W., Lewin, W., Sztajno, M., Tr{\"u}mper, J. 1988,
  A\&A, 192, 147

\bibitem[\protect\astroncite{van Paradijs et~al.}{1986}]{psl+86}
van Paradijs, J., Sztajno, M., Lewin, W., et~al. 1986, MNRAS, 221, 617

\bibitem[\protect\astroncite{van Paradijs \& White}{1995}]{pw95}
van Paradijs, J., White, N. 1995, ApJ (Letters), 447, L33

\end{thebibliography}
\end{document}